# 3D Reconstruction Using Optical Images

**Edward N. Tsyganov, Pietro P. Antich, Ralph P. Mason , Robert W. Parkey, Serguei Y. Seliounine, Nikolai V. Slavine, Alexander I. Zinchenko**

The University of Texas Southwestern Medical Center in Dallas

Department of Radiology, Advanced Radiological Sciences

5323 Harry Hines Blvd., Dallas, Texas 75390-9071

Abstract

Various imaging methods for small animals are rapidly gaining acceptance in biology and medical research. Optical imaging is a very fast and convenient method of biological interrogation, but suffers from significant disadvantages, such as the absence of 3D image reconstruction algorithms. This have up until now impeded progress to a quantitative stage. Here we propose a 3D reconstruction algorithm for imaging light emitted at depths in small living animals. We believe the methods discussed here are novel and lead to a novel imaging paradigm, which we call Light Emission Tomography.

## Introduction

Optical imaging is a powerful tool in biology and medical research. Optical imaging uses bioluminescence of injected luciferin, or fluorescence of fluorophore or labeled biomolecule with an external light source.[1, 2] Cooled CCD permit the acquisition of images at very low intensity level in a wide wavelength range and optical imaging is a fast and convenient method of biological interrogation.

Among the disadvantages of optical methods are the strong scattering and absorption effects in tissue and complexity of light transport, resulting in the absence of a practical, high-resolution 3D image reconstruction algorithm. Some initial attempts of depth determination have been made [3] but remain to be put into practice. Hence, optical imaging has not yet reached a full quantitative stage.

Here we propose a 3D reconstruction algorithm for small animal imaging. We believe our approach is original one and different from other proposed approaches. We also discuss ways to take into account scattering and absorption corrections to make possible a quantitative description of biological processes in optical imaging. The first formulation of the contents of this paper is in [4].

## Task Formulation

Although one can perceive a 3D image from photographs using stereo projections, this requires a certain amount of human brain processing.

The crucial requirement in this type of processing is the identification of the same parts of an object presented in both photographs. Bubble chamber experiments in high-energy physics are an example of stereo measurements. In this example, an observer has a sharp detail to act upon, such as a vertex of a particle interaction, presented in both photographs.[5] This method, however, is not practical for use in optical medical imaging. Instead, we propose to use a mathematically defined algorithm to do all of the 3D processing, without any observer.

We adopted an approach that is closely related to 3D processing in other medical imaging modalities. The general method to produce a 3D image from medical imaging scans is a backprojection of measured intensities into the space occupied by the object. In x-ray Computed Tomography (CT) the x-ray source and a detector position is used to define a backprojection line. Single Photon Emission Computed Tomography (SPECT) uses the position of a gamma interaction and a directional collimator to backproject lines. In Positron Emission Tomography (PET) the line, connecting the coordinates of the interactions of two co-linear gamma rays from positron annihilation defines the backprojection.

In optical imaging, we must first solve the problem of backprojection through the optical objective. An optical objective is a complex optical system, and the backprojection task does not appear to be straightforward, even for the simplest case (see Figure 1). Even if the effective position and the focus of the objective are known, a problem remains: the object O is 3-dimensional, and the image I by definition has only two dimensions.

To solve the problem, we first decided to use for backprojection only the central line in Figure 1. In this approximation, the center of the objective acts as a pinhole collimator in SPECT. Although the central line differs slightly from the real paths of the all photons, it pinpoints the same position on the object. The backprojection line is produced when the center of a pixel at the image is connected to the effective center of the objective. After this procedure, the powerful methodology of medical imaging can be applied for 3D reconstruction of the image.

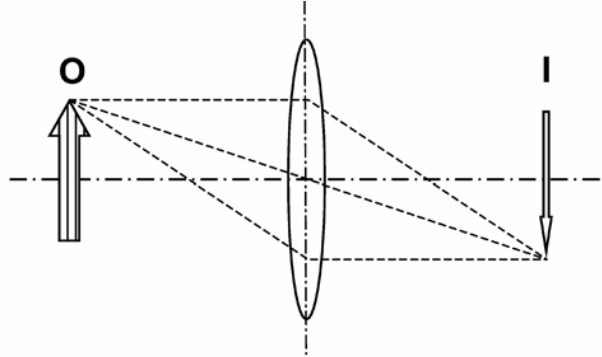

Figure 1. Simplified schematics of optical imaging. O – the object, I – the image.

## *Experiment*

To test this approach, we considered a simple experimental setup and proceeded as follows. The setup for experimental tests of the method is presented in Figure 2. The object was simulated by two light emitting diodes. A simple digital camera, Sony DSC-P51, was used to image projections. The camera had an 8-bit 1600x1200 pixel CCD with 3.275 μm pixel sizes; the focal distance was 6.3-12.6 mm. The camera was mounted on a simple goniometer. Figure 3 demonstrates a raw image of the object.

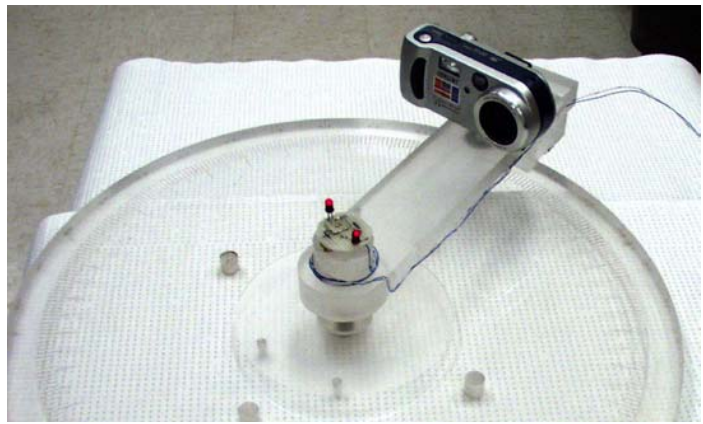

Figure 2. Setup for experimental tests of the method.

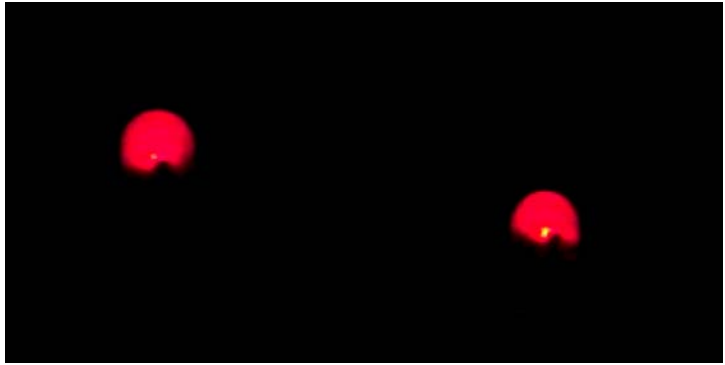

Figure 3. A photograph of the objects at an angle of $0^o$.

Twelve angular views with a step of $30^o$ were acquired to produce the tomographic 3D image of the objects. We estimate an angular position error less than $0.1^o$ (i.e., about 0.3 mm in the image space).

## Image reconstruction

To simplify the reconstruction task, we converted the raw images to a gray scale. Then, we connected pixel centers to the center of objective and backprojected the lines into the object volume. We took the number of such lines to be proportional to the light intensity detected by the pixel. To reconstruct the 3D image we use the Maximum Likelihood Expectation Maximization (ML-EM) method [6] given by:

$$n_j^{k+1} = \frac{n_j^k}{\sum_{i=1}^{I} a_{ij}} \sum_{i=1}^{M} a_{ij} \frac{1}{q_i^k} \quad \text{with} \quad q_i^k = \sum_{j=1}^{J} a_{ij} n_j^k \qquad (1)$$

Here $q_i^k$ is the expected count in line of response (LOR) i if the voxel intensity is $n_i^k$ (at the k-th iteration), $a_{ij}$ is the probability of an emission from voxel j being detected along LOR i, M is the number of measured events and I is the number of all possible system LOR's. J is the number of voxels in the image.

Figure 4 presents a projection of the 3D reconstruction using the maximum intensity voxel projection technique (the voxel size is 0.25x0.25x0.25 mm$^3$). This image is taken for a 0$^o$ rotation.

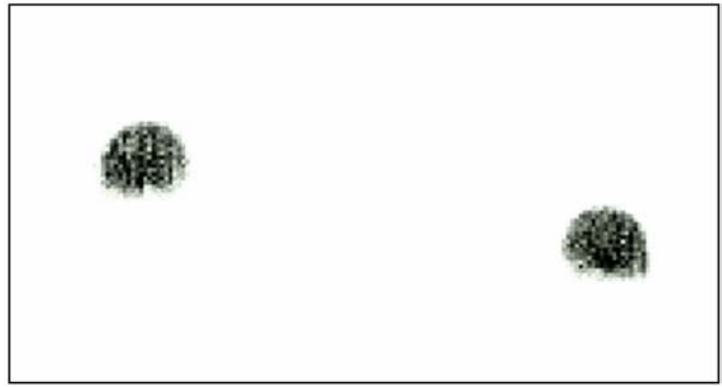

Figure 4. 2D projection of 3D image of the light emitting diodes. Voxel size is 0.25 mm.

Figure 5 shows 0.5 mm thick slice of the image in the X-Z plane.

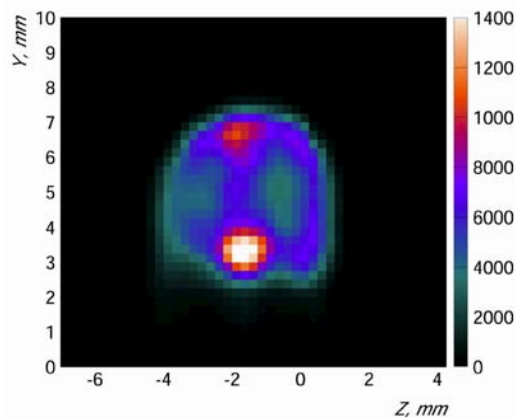

Figure 5. 0.5 mm slice of the LED image in the vertical plane.

Figure 6 presents a photograph and a maximal intensity voxel projection of a light emitting diode in detail.

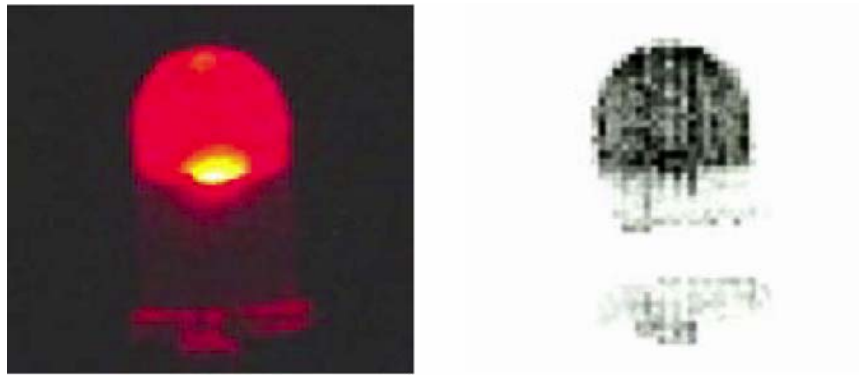

Figure 6. Left – photograph of the light emitting diode, right – the maximum intensity voxel projection of the 3D image. Rotation of 3D image reveals the diode's inner structures.

Figure 7 shows ML-EM 3D reconstruction of a sculptural portrait of a famous composer. This Figure shows that the method works not only for emitted, but also for reflected (diffused) light.

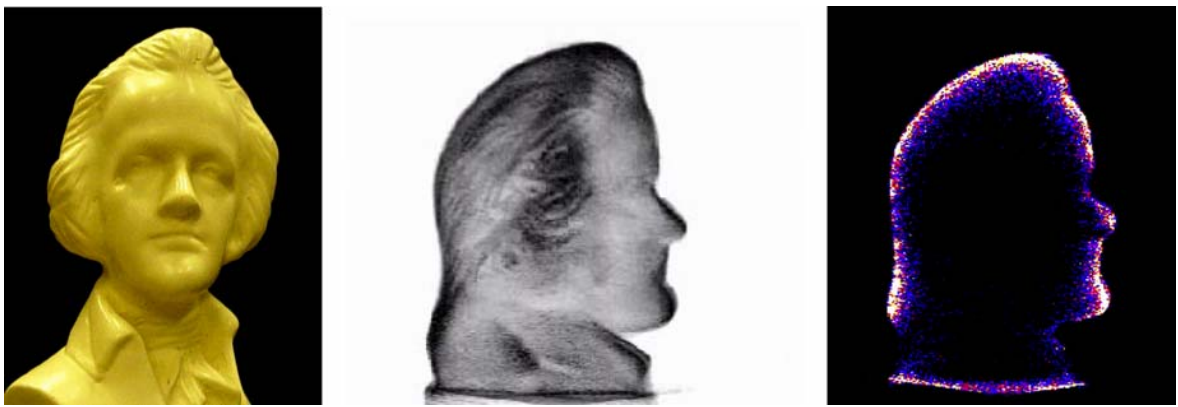

Figure 7. 3D image of a sculpture. Left – a photograph, middle – maximum intensity voxel projection of 3D image, right –0.25 mm central slice of 3D image.

Figure 8 compares a photograph of a nude mouse taken in a micro SPECT scanner and a maximum intensity voxel projection of the reconstructed 3D image. Obviously, a 3D image of a mouse surface is a good complementary modality to a micro SPECT.

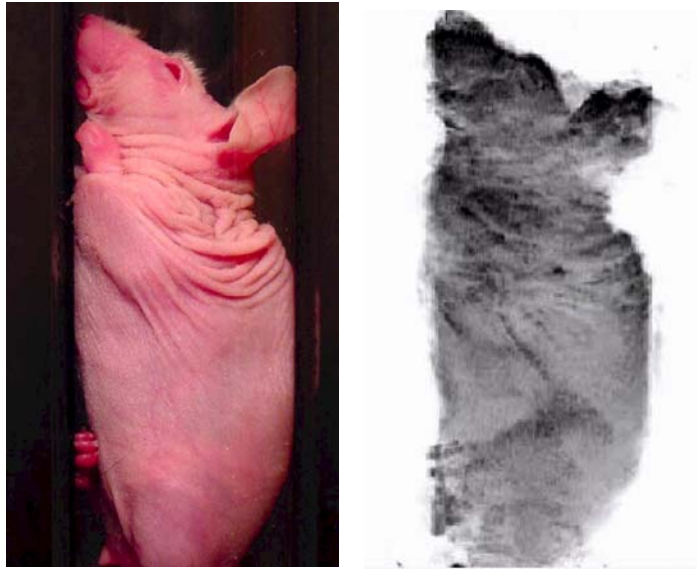

Figure 8. Left: a photograph of a nude mouse in a micro-SPECT scanner, right: a maximum intensity voxel projection of the reconstructed optical 3D image.

## *Light Diffusion and Absorption*

Because of the high absorption of green and yellow light in hemoglobin, it has been proposed to use red or near-infrared light for optical imaging, i.e., the wavelength region between 600 and 800 nm using red fluorescence proteins (RFP). However, even with low absorption red-infrared light, scattering processes present the real difficulty in optical imaging. For example, the scattering attenuation coefficient in muscle tissue is 21 cm$^{-1}$ for 670 nm light,[7] nearly 200 times that for absorption.

The main scattering mechanism is light refraction due to spatial variations in the refractive indices of a tissue medium. Monte Carlo simulation of photon scattering in biological tissues has been successful.[8] Following this direction, we intend to introduce scattering corrections into small animal optical images similar to the deconvolution approach adopted for PET scattering corrections and system modeling in [9].

The scattering phantom was a truncated cone of wax with light emitting diodes placed inside two holes (Fig. 8). ML-EM with no deconvolution reconstructs a distribution confined essentially to the outer surface of the phantom.

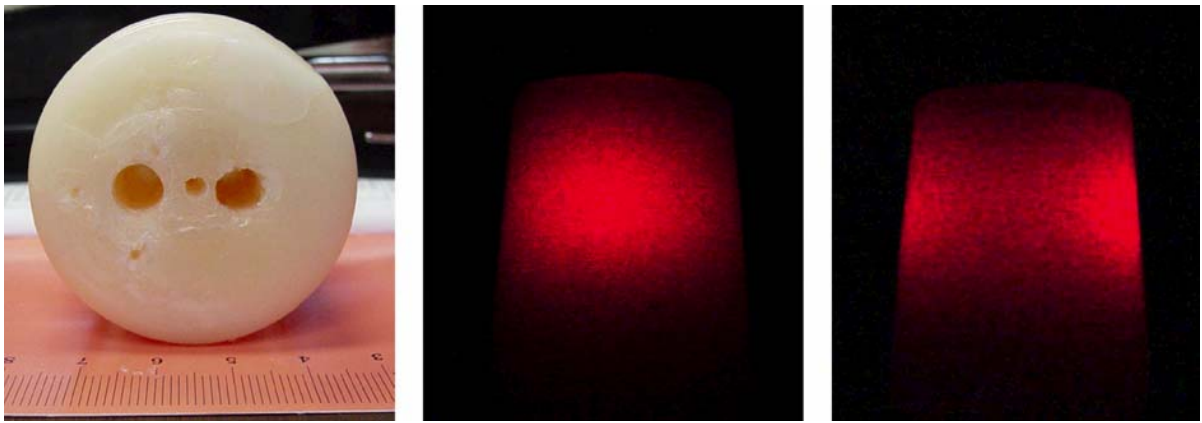

Figure 9. Left – wax phantom with two holes for LEDs. Right – two orthogonal views of the phantom in dark.

Figure 9, (left panel) presents a 0.5 mm slice of the reconstructed image already shown. Figure 9, middle, presents the results of our first attempt at deconvolving light scattering assuming a simple shift-invariant scattering kernel. This is patently a false assumption, which, however, would serve to indicate further avenues for progress. The slice is again 0.5 mm thick. Figure 9, right, is a superposition of two images.

Thus, this first oversimplified attempt is remarkably successful taking into account very crude simplification of the light transport process. More correct modeling is now in progress.

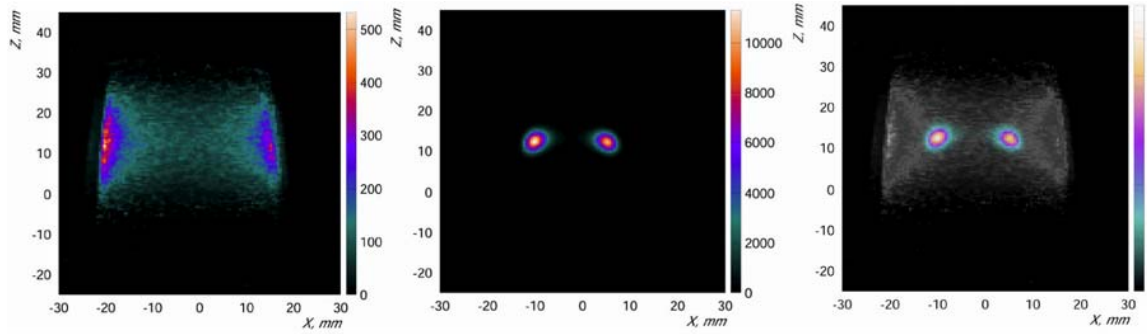

Figure 10. Wax phantom reconstructed. Left – with no scattering treatment, middle – with scattering deconvolution, kernel 8 mm, and right – superposition of the left and the middle images.

## *Conclusion*

We have outlined a possible algorithm for 3D optical image reconstruction. While the algorithm works well in the no-scattering case, scattering corrections are expected to be large in a realistic tissue media case. Although the deblurring mechanism looks straightforward, its computational effectiveness should be optimized and is a topic of our future work.

## *Acknowledgements*

We are thankful to Dr. G. Arbique, Dr. M. Lewis, and T. Soesbe for useful discussions and help, and Dr. R. McColl for help in computing. The work was supported in part by the Cancer Imaging Program (an NCI Pre-ICMIC) 1P 20.

*References*

1. Christoph Bremer, Sebastian Bredow, Umar Mahmood, Ralph Weissleder, Ching-Hsuan Tung, "Optical Imaging of Matrix Metalloproteinase-2 Activity in Tumors: Feasibility Study in a Mouse Model", *Radiology* 2001; 221:523-529.

2. Christopher H. Contag and Michael H. Bachmann, "Advances in Vivo Bioluminescence Imaging of Gene Expression", *Annu. Rev. Biomed. Eng.* 2002, 4:235-260.

3. Oliver Coguos, Tamara L. Troy, Dragana Jekic-MsMullen, Bradley W. Rice, "Determination of depth of *in-vivo* bioluminescent signals using spectral imaging techniques", *Proceedings of SPIE*, v. 4967, 2003, p. 37-45.

4. Edward N Tsyganov, Pietro P. Antich, Robert W. Parkey, Serguei Y. Seliounine, Nikolai V. Slavine, "3D Reconstruction In Optical Imaging", Internal Report of The University of Texas Southwestern Medical Center, May 9, 2003.

5. D. A. Glaser. "Elementary Particles and Bubble Chambers", in Nobel *Lecture Physics, 1942-1962,* Elsevier, Amsterdam, 1964, p. 529.

6. L. A. Shepp and Y. Vardi, "Maximum Likelihood Reconstruction for Emission Tomography", *IEEE Transactions on Medical Imaging,* v. MI-12, No. 2, p. 113-121, 1982.

7. R. Srinivasan, D. Kumar and Megha Singh, "Optical Tissue-Equivalent Phantoms for Medical Imaging", *Trends Biomater. Artif. Organs,* v. 15(2) pp 42-47, 2002.

8. Wilson B.C. and Adam G., "A Monte Carlo model for the absorption and flux distributions of light in tissues", *Med. Phys.,* 10, pp 824-830, 198

9. Roido Manavaki, Andrew J. Reader, Claudia Keller, John Missimer and Richard J. Walledge, "Scatter Modeling for 3-D PET List-Mode EM Reconstruction", *in Record of IEEE Nuclear Science Symposium and Medical Imaging Conference 2002,* Norfolk, Virginia, November 11-16, 2002; Andrew J. Reader et al., "One-Pass List-Mode EM Algorithm for High Resolution 3D PET Image Reconstruction into Large Arrays", *IEEE Transactions on Nuclear Science,* v.49, no 3, June 2002, p. 693.

*Figure captions*

1. Simplified schematics of optical imaging. O – the object, I – the image.
2. Setup for experimental tests of the method.
3. A photograph of the objects at an angle of $0^o$.
4. 2D projection of 3D image of the light emitting diodes. Voxel size is 0.25 mm. *Supplementary data: thir-025.mpg, 838 KB.*
5. 0.5 mm slice of the LED image in the vertical plane.
6. Left – photograph of the light emitting diode, right – the maximum intensity voxel projection of the 3D image. Rotation of 3D image reveals the diode's inner structures. *Supplementary data: phot5-v0125.mpg, 2,035 KB.*
7. 3D image of a sculpture. Left – a photograph, middle – maximum intensity voxel projection of 3D image, right –0.25 mm central slice of 3D image. *Supplementary data: moz025-mod OK.mpg, 2,384 KB.*
8. Left: a photograph of a nude mouse in a micro-SPECT scanner, right: a maximum intensity voxel projection of the reconstructed optical 3D image. *Supplementary data: nudm-v025-cut.mpg, 3,453 KB.*
9. Left – wax phantom with two holes for LEDs. Right – two orthogonal views of the phantom in dark.
10. Wax phantom reconstructed. Left – with no scattering treatment, middle – with scattering deconvolution, kernel 8 mm, and right – superposition of the left and the middle images.